\documentclass[12pt]{article}
\begin{document}
\def\ti#1{{}_{#1}}
\def\up#1#2{\stackrel{\ti{(#1)}}{#2}}
\def\th{\theta}
\def\vecr#1{\mid{#1}\rangle}
\def\vecl#1{\langle{#1}\mid}
\begin{center}
{\large\bf Polarization scattering by soliton-soliton collisions}\\[.5cm]
{Valery Shchesnovich}\\
Division for Optical Problems in Information
Technologies, Academy of Sciences of Belarus, Zhodinskaya
Str. 1/2, 220141 Minsk, Republic of Belarus.
\end{center}
\medskip

\begin{center}
\parbox{0.8\textwidth}
{\small{\bf Abstract.} Collision of two solitons of the Manakov system is
studied analytically. Existence of a complete polarization mode switching
regime is proved and the parameters of solitons prepared for polarization
switching are found. }
\end{center}
\medskip

The idea of an analytical approach to the description of polarized (i.e.,
vector) solitons in nonlinear optical fibers is based on the
fact~\cite{Menyuk} that for some values of the parameters the original system
of coupled nonlinear Schr{\"o}dinger equations transforms to the Manakov
system

$$
i{q_1}_z+\frac{1}{2}{q_1}_{\tau\tau}+\left(|q_1|^2+|q_2|^2\right)q_1 =0, $$
\vspace{-7mm}
$$
\eqno(1)
$$
\vspace{-7mm}
$$ i{q_2}_z+\frac{1}{2}{q_2}_{\tau\tau}+\left(|q_2|^2+|q_1|^2\right)q_2 =0. $$
This system  was shown to be integrable by Manakov~\cite{Manakov}. Here $q_1$
and $q_2$ are normalized envelopes of the two modes of the soliton, while
$\tau$ and $z$ are, correspondingly, the normalized retarded time and
distance along the fiber. The one-soliton solution for Eq.~(1) reads
 $$
q_l=-2i\eta\theta_l\,\frac{\exp\left(-2i[\xi\tau+2(\xi^2-\eta^2)z]\right)}
{{\rm cosh}\left(2\eta[\tau-\tau_0+2\xi z]\right)},\quad l=1,2, \eqno(2)
 $$
where $\theta_l=|\theta_l|e^{i\phi_l}$, $l=1,2$, $|\theta_l|$ and~$\phi_l$
are the intensity and initial phase of the $l$-th polarization component of
the soliton, $|\theta_1|^2+|\theta_2|^2=1$, and $2\xi$ and~$2\eta$ are the
velocity and amplitude of the soliton, correspondingly. Many-soliton
solutions of the Manakov system can be easily retrieved with the help of the
Riemann-Hilbert problem approach. A detailed analysis of the associated
Riemann-Hilbert problem and the perturbation theory for Manakov solitons can
be found in Ref.~[3]. The general soliton matrices\footnote{Discovered quite
recently.}, which allow for the complete classification of all possible
soliton solutions to the Manakov system as well, can be found in
Ref~\cite{JMP}.

Below we will give the formulas describing collision of two Manakov solitons
with arbitrary polarization states. To conveniently describe the phase and
polarization state of the two-component soliton let us introduce a
two-dimensional column vector $\vecr{\theta}$, which will be referred to
below as the theta-parameter. The Hermitian-transposed row vector will be
denoted as $\vecl{\theta}$. Analysis of the two-soliton solution of the
Manakov system~(1) under the condition that $\xi_2>\xi_1$ leads to the
following transformation formulas for the parameters $\vecr{\theta^{(1)}}$
and~$\vecr{\theta^{(2)}}$:
$$
\vecr{\up{+}{\theta}^{(1)}}=\exp\biggl(i{\rm arg}
\biggr(\frac{k_1-\overline k_2}{k_1-k_2}\biggl)-\beta\biggr)\left({\rm I}
+\frac{k_2-\overline k_2}{k_1-k_2}\vecr{\up{-}{\theta}^{(2)}}
\vecl{\up{-}{\theta}^{(2)}}\right)\vecr{\up{-}{\theta}^{(1)}},
$$
$$
\eqno(3)
$$
$$
\vecr{\up{+}{\theta}^{(2)}}=\exp\biggl(i{\rm arg}\biggr(\frac{k_1-\bar
k_2}{\overline k_1-\overline k_2}\biggl)-\beta\biggr)\left({\rm I}+\frac{k_1-\bar
k_1}{\overline k_1-\overline k_2}\vecr{\up{-}{\theta}^{(1)}}\vecl{\up{-}{\theta}^{(1)}}\right)
\vecr{\up{-}{\theta}^{(2)}},
$$
where the top signs, minus and plus, denote the theta-parameters before and
after the collision, correspondingly, $k_j=\xi_j+\eta_j$, $j=1,2$, and
$\beta$ reads
$$
\beta=\frac{1}{2}{\rm ln}\left(1-\frac{(k_2-\overline k_2)(k_1-\overline
k_1)}
{|k_1-k_2|^2}\bigg|\langle\up{-}{\theta}^{(1)}\vecr{\up{-}{\theta}^{(2)}}\bigg|^2\right).
\eqno(4)$$
 The  transformation given by Eqs.~(3) and (4)  was first derived by Manakov
 \cite{Manakov} in an equivalent form.

 It is easy to verify that the theta-parameters satisfy  the following simple  identity
$$
\langle\up{+}{\theta}^{(1)}\vecr{\up{+}{\theta}^{(2)}}=\exp\left(4i\,{\rm
arg}(k_1-k_2)\right)\langle\up{-}{\theta}^{(1)}\vecr{\up{-}{\theta}^{(2)}}.
\eqno(5)$$
 Additionally to  transformation~(3)-(4), the interaction of two Manakov solitons leads
to the finite shifts in the time of arrival parameters of the two solitons:
$$
\Delta\tau_o^{(1)}\equiv\up{+}{\tau}_o^{(1)}-\up{-}{\tau}_o^{(1)}
=\frac{1}{2\eta_1}\left({\rm ln}\bigg|\frac{k_1-\overline k_2}{k_1-k_2}\bigg|+\beta\right),
$$
$$
\eqno(6)
$$
$$
\Delta\tau_o^{(2)}\equiv\up{+}{\tau}_o^{(2)}-\up{-}{\tau}_o^{(2)}
=-\frac{1}{2\eta_2}\left({\rm ln}\bigg|\frac{k_1-\overline k_2}{k_1-k_2}\bigg|+\beta\right).
$$
Equations (3), (4), and~(6) completely describe the collision of two Manakov
solitons.

The transformation given by Eqs.~(3)-(4) means that the colliding solitons
significantly alter each other's polarizations and phases. This fact was
already mentioned by Manakov \cite{Manakov}. Moreover, there are the
polarization switching regimes of collision, i.e. when we have
$\up{+}{\theta}^{(n)}_l=0$ but $\up{-}{\theta}^{(n)}_l\ne 0$, for some $n$
and $l$ ($n=2$ and $l=2$ below). The polarization switching regime was
discovered in Ref.~\cite{RLH} in the numerical simulations of the general
two-soliton solution. To  analytically prove the existence of such a regime
let us invert one of the equations from system (3) for the theta-parameters
of the solitons. For instance, given $\vecr{\up{+}{\theta}^{(2)}}$
and~$\vecr{\up{-}{\theta}^{(1)}}$ we obtain for the rest of the
theta-parameters\footnote{Another way to derive Eqs.~(7)-(8) is to use the
asymptotic expansion of general two-soliton solution on the two
characteristics, defined by $\tau + 2\xi_n z = \mathrm{const.}$, $n=1,2$, and
the fact that there are exactly two equivalent factored representations of
the soliton matrix: $\Gamma(k) = \chi_1(k)\chi_2(k) =
\widetilde{\chi_2}(k)\widetilde{\chi_1}(k)$, for notations consult
Ref.~\cite{JMP}.}:
 $$
\vecr{\up{-}{\theta}^{(2)}}=\exp\biggl(i{\rm arg}
\biggr(\frac{\overline k_1-k_2}{k_1-k_2}\biggl)+\beta\biggr)\left({\rm I}
+\frac{k_1-\overline k_1}{\overline k_2-k_1}\vecr{\up{-}{\theta}^{(1)}}
\vecl{\up{-}{\theta}^{(1)}}\right)\vecr{\up{+}{\theta}^{(2)}},
 $$
 $$
\eqno(7)
 $$
 $$
\vecr{\up{+}{\theta}^{(1)}}=\exp\biggl(i{\rm arg}\biggr(\frac{k_1-\bar k_2}{
k_1- k_2}\biggl)+\beta\biggr)\left({\rm I}+\frac{k_2-\bar k_2}{\overline
k_1-k_2}\vecr{\up{+}{\theta}^{(2)}}\vecl{\up{+}{\theta}^{(2)}}\right)
\vecr{\up{-}{\theta}^{(1)}}.
 $$
Importantly, the set of the theta-parameters $\{\vecr{\up{-}{\theta}^{(1)}},
\vecr{\up{+}{\theta}^{(2)}}\}$ is a close one due to the fact that $\beta$
admits also the following equivalent representation
 $$
\beta=-\frac{1}{2}{\rm ln}\left(1+\frac{(k_2-\overline k_2)(k_1-\overline
k_1)} {|k_1-\overline
k_2|^2}\bigg|\langle\up{-}{\theta}^{(1)}\vecr{\up{+}{\theta}^{(2)}}\bigg|^2\right).
 \eqno(8)$$
 Let us put
 $$
\vecr{\up{-}{\theta}^{(1)}}=\left(\begin{array}{c}{\rm cos}\alpha_1 e^{i\phi_1}\\
{\rm sin}\alpha_1 e^{i\phi_2}\end{array}\right),\quad
\vecr{\up{+}{\theta}^{(2)}}=\left(\begin{array}{c}e^{i\psi}\\0\end{array}\right),
\eqno(9)
 $$
where $\alpha_1$, $\phi_1$, $\phi_2$ and~$\psi$ are given. Define also
$$
\vecr{\up{-}{\theta}^{(2)}}=\left(\begin{array}{c}{\rm cos}\alpha_2 e^{i\psi_1}\\
{\rm sin}\alpha_2 e^{i\psi_2}\end{array}\right). \eqno(10)
 $$
For the initial (i.e. before the collision) theta-parameter of soliton~2 we
obtain from Eq.~(7)
$$
\vecr{\up{-}{\theta}^{(2)}}=\frac{{\rm exp}\left(i{\rm arg}(\frac{k_2-\overline k_1}
{k_2-k_1})+i\psi\right)}{\left(1-\frac{4\eta_1\eta_2}{|\overline k_2-k_1|^2}
{\rm cos}^2\alpha_1\right)^{\frac{1}{2}}}
\left(\begin{array}{c}1+\frac{2i\eta_1}{\overline k_2-k_1}{\rm cos}^2\alpha_1\\
\frac{2i\eta_1}{\overline k_2-k_1}{\rm cos}\alpha_1{\rm sin}\alpha_1
e^{(i\phi_2-i\phi_1)}\end{array}\right) \eqno(11)$$
 One can use Eqs.~(10) and (11) to find the values of the
polarization angle and  phase difference of soliton~2 prepared for the
complete polarization switching due to collision with soliton~1 with the
phase vector fixed by Eq.~(9):
$$
\alpha_2=-{\rm arctan}\left(\frac{\eta_1{\rm sin}(2\alpha_1)}
{\biggl((\xi_2-\xi_1)^2+\Bigl(\eta_2-\eta_1{\rm cos}(2\alpha_1)\Bigr)^2
\biggr)^{\frac{1}{2}}}\right),
$$
$$
\eqno(12)
$$
$$
\psi_2-\psi_1=\phi_2-\phi_1+\frac{\pi}{2}+{\rm arctan}
\left(\frac{\eta_2-\eta_1{\rm cos}(2\alpha_1)}{\xi_2-\xi_1}\right).
$$

Therefore, it is analitically shown that vector solitons governed by the
system of coupled nonlinear Schr{\"o}dinger equations exhibit a novel type of
soliton collision -- the soliton polarization state switching. In the case
when the Manakov system is the relevant model the polarization switching will
be complete.  The soliton polarization state switching can be used for the
logic elements based on vector solitons.

\vskip 1cm \noindent{\large\bf Acknowledgement}

The work was supported by the Fundamental Research Foundation of Belarus
under contract No. M$\Pi$96-06. The author is grateful to Dr. Takayuki
Tsuchida for pointing out a misprint.

\end{document}